\documentclass[11pt, a4paper,english,superscriptaddress,aps,nofootinbib]{revtex4}
\usepackage{amsmath,amssymb,graphicx}
\makeatletter
\usepackage{babel}
\usepackage[active]{srcltx}
\usepackage{graphicx,color}
\usepackage{changebar}
\usepackage{hyperref}
\hypersetup{
	colorlinks=true,
	urlcolor= blue,
	citecolor=blue,
	linkcolor= blue,
	bookmarks=true,
	bookmarksopen=false,
}
\usepackage[T1]{fontenc}
\usepackage{ae}
\usepackage[ansinew]{inputenc}
\usepackage{amsmath}
\usepackage{braket}
\usepackage{mathtools}
\usepackage{slashed}
\usepackage{empheq}
\usepackage{tikz}
\usepackage{multirow}
\usetikzlibrary{decorations.pathmorphing}
\usepackage[caption = false]{subfig}

\usepackage{graphicx}
\usepackage{amsfonts}
\usepackage{amsmath}
\newcommand{\be}{\begin{equation}}
\newcommand{\ee}{\end{equation}}
\newcommand{\bea}{\begin{eqnarray}}
\newcommand{\eea}{\end{eqnarray}}

\begin{document}
\title{First-order perturbations of G\"{o}del-type metrics in non-dynamical\\
Chern-Simons modified gravity}

\author{B. Altschul}
\email[]{baltschu@physics.sc.edu}
\affiliation{Department of Physics and Astronomy, University of South Carolina,\\
Columbia, South Carolina 29208, USA}

\author{J. R. Nascimento}
\email[]{jroberto@fisica.ufpb.br}
\affiliation{Departamento de F\'{\i}sica, Universidade Federal da 
Para\'{\i}ba,\\
 Caixa Postal 5008, 58051-970, Jo\~ao Pessoa, Para\'{\i}ba, Brazil}

\author{A. Yu. Petrov}
\email[]{petrov@fisica.ufpb.br}
\affiliation{Departamento de F\'{\i}sica, Universidade Federal da Para\'{\i}ba,\\
 Caixa Postal 5008, 58051-970, Jo\~ao Pessoa, Para\'{\i}ba, Brazil}

\author{P. J. Porf\'{\i}rio}
\email[]{pporfirio@fisica.ufpb.br}
\affiliation{Departamento de F\'{\i}sica, Universidade Federal da 
Para\'{\i}ba,\\
 Caixa Postal 5008, 58051-970, Jo\~ao Pessoa, Para\'{\i}ba, Brazil}

\begin{abstract}
G\"{o}del-type metrics that are homogeneous in both space and time remain, like the Schwarzschild metric, consistent within Chern-Simons modified gravity; this is true in both the non-dynamical and dynamical frameworks, each of which involves an additional pseudoscalar field coupled to the Pontryagin density. In this paper, we consider stationary first-order perturbations to these metrics in the non-dynamical framework. Under certain assumptions we find analytical solutions to the perturbed field equations. The solutions of the first-order field equations break the translational and cylindrical symmetries of the unperturbed metrics. The effective potential controlling planar geodesic orbits is also affected by the perturbation parameter, which changes the equilibrium radii for the orbits of both massive particles and massless photons.

\end{abstract}

\maketitle
 
 \section{INTRODUCTION}
 
The exploration of alternative theories of gravity has drawn great attention in recent decades, driven in part by  recent cosmological observations \cite{Riess, Riess1}. To explain the phenomena of the late-time accelerated expansion of the Universe and the rotation curves of galaxies in clusters within the General Relativity (GR) framework, it is necessary to postulate the existence of matter and energy sources (dark matter and dark energy) of unknown nature, which fill the whole Universe and are supposed to dominate the gravitational dynamics at cosmic scales. The difficulties with achieving direct detection of these dark components and distinguishing their effects from modifications of GR at large scales motivate the exploration of alternative theories of gravitation. On the other hand, at small scales, it is well known that GR is non-renormalizable \cite{tHooft:1974toh}, which requires that GR be subject to some unknown kind of completion in the ultraviolet regime.

We shall give particular attention to Chern-Simons (CS) modified gravity \cite{Jackiw:2003pm, Alexander:2009tp} in this work. This modification in this theory consists of adding to the Einstein-Hilbert action a Pontryagin term coupled to a pseudoscalar field (which, in general, may be either dynamical or non-dynamical). The Pontryagin term commonly appears in a variety of distinct contexts of high-energy physics---in field theory, where it is proportional to the gravitational anomaly \cite{AlvarezGaume:1983ig}; in loop quantum gravity \cite{Ashtekar:1988sw}; in string theory, in which it emerges via the Green-Schwarz mechanism responsible for the anomaly cancellation in $SO(32)$ and $E_8\times E_8$ heterotic string theories \cite{Green:1984sg}; within studies of local Lorentz symmetry breaking \cite{Kostelecky:2003fs,Bluhm:2014oua}; and in other contexts (see, for example, \cite{Mag} and references therein). Such multifarious occurrences demonstrate agreement on the highly relevant role of the Pontryagin density in high-energy physics; and it follows that CS-modified gravity is not just one of many arbitrary modifications of GR, but is actually closely tied to concepts that have many nontrivial applications.  

Major interest in CS gravity was first inspired by the paper \cite{Witten:1988hc}, which called attention to it in three-dimensional CS gravity. The concept of a topological gravity theory was later generalized \cite{Chamseddine:1989nu}  to higher odd dimensions, and in \cite{Chamseddine:1990gk}, this was further generalized, via the introduction of an additional spinless field, to even-dimensional spacetimes. Since then, various interesting issues in CS gravity theories have been explored; see, for example, \cite{HasZan} and references therein. Among the specific motivations for studying CS gravity models, their application, in eleven-dimensional spacetimes, to M-theory is worth mentioning as well \cite{Izaurieta:2006aj}.

Special attention has been paid to the four-dimensional CS modified gravity, because of its relation to the problems of Lorentz and CPT symmetry breaking. The original study emphasizing this relationship was presented in the paper by Jackiw and Pi \cite{Jackiw:2003pm}; this paper considered a setup in which a pseudoscalar field $\phi$ coupled to the Pontryagin density has no dynamics; that is, $\phi$ is an external prescribed quantity or coefficient. This formulation is known as the non-dynamical CS-modified gravity (NDCSMG) theory. To understand the main motivation for the study of NDCSMG, it is necessary to interpret it in the context of local Lorentz and diffeomorphism symmetry breaking. For that, we cite the Standard Model Extended (SME) \cite{Kostelecky:2003fs}, which is an effective field theory incorporating all possible coefficients for (local) Lorentz, CPT, and diffeomorphism violation coupled to particle and gravitational sources. In particular, NDCSMG is one specific and unusually interesting CPT-violating term, as pointed out in \cite{Kostelecky:2003fs}. If the local Lorentz and diffeomorphism symmetry breaking stems from some dynamical mechanism occurring at the Planck scale \cite{Kostelecky:1988zi, Kostelecky:1989jp}, then the natural way to incorporate such violation into a low-energy effective theory of gravity is to introduce tensor background fields (spacetime anisotropies) that break the spacetime symmetries either spontaneously or explicitly. In particular, the symmetries invovled are explicitly broken in NDCSMG \cite{Bluhm:2014oua}, while they are spontaneously broken in dynamical CS-modified gravity (DCSMG)---in which $\phi$ is promoted to being a fully dynamical pseudoscalar field with its own interactions \cite{Alexander:2009tp}.

As well as being a useful model of explicit gravitational CPT symmetry violation, NDCSMG provides an important first step toward understanding DCSMG, since many solutions within NDCSMG remain solutions of the equations of motion for DCSMG; see for example \cite{Porfirio:2016nzr, Porfirio}. Indeed, the first known solutions in CS modified gravity, which inspired further studies of the theory, were obtained in the specifically non-dynamical case \cite{Jackiw:2003pm,Konno:2007ze}. In the intervening years, there have been interesting studies of both versions of the theory. One may remark, for example, on the context of black holes---where it has been shown that the Schwarzschild metric~\cite{Jackiw:2003pm} remains a solution of the gravitational field equations within the model, while new solutions describing slowly rotating black holes, which have no analogues in conventional GR, have also been found \cite{Konno:2007ze, Yunes:2009hc}. More recently, with the experimental observation of gravitational waves, many works have searched for parity-violating gravitational radiation signatures within the CS-modified gravity framework \cite{Bartolo:2017szm, Bordin:2020eui}.

One special class of metrics that has historically attracted a great deal of attention in GR~\cite{Reb}, as well as in alternative geometric theories of gravity, are the G\"{o}del-type metrics \cite{ourgodel, Porfirio:2016nzr, Porfirio}. The richness of their physical properties alone is a sufficient motivation to consider them. One of the most striking features of this class of metrics is the possibility for the existence of closed timelike curves (CTCs), which allow observers to travel back to their own pasts---thus breaking the causality principle and violating the chronology protection conjecture \cite{Hawking:1991nk} which many of our physical theories are based on --- even without destroying the {\em local} Lorentzian causality properties of GR. Obviously, a particularly interesting example of a G\"{o}del-type metric is the well-known G\"{o}del spacetime that was historically one of the first exact solutions of GR that displayed CTCs. However, there are also other examples of spacetime metrics exhibiting CTCs, such as the Van Stockum metric \cite{VanS} and Gott time machines \cite{Gott}, which could be studied in similar fashions. Yet even within G\"{o}del-type metrics possessing CTCs there is are completely causal regions (where CTCs are entirely avoidable), depending on the relations amongst the various metric parameters. The consistency of G\"{o}del-type metrics has been verified in both formulations of CS-modified gravity \cite{Porfirio:2016nzr, Porfirio}, and it is also worth calling attention to the studies related to the quasi-normal modes in the context of G\"{o}del-type metrics in \cite{Konoplya:2011ag, Konoplya:2011hf, Konoplya:2011it}.

In this work, we shall concentrate our efforts on an analysis of time-independent first-order perturbations around G\"{o}del-type metrics within NDCSMG, paying special attention to the possible breaking of the Lorentz symmetry due to the perturbations. This will entail taking as the undeformed background solutions the ones previously found in \cite{Porfirio:2016nzr}. Since this is a primarily exploratory, theoretical exercise, it makes sense to limit our attention to time-independent perturbations, even though time-dependent perturbations around smooth, simple background solutions are obviously extremely important in practical cosmology. However, despite G\"{o}del-type metrics and their deformations not being good candidates to describe our Universe (because they include regions with CTCs), they are nonetheless of great theoretical interest because of their peculiar global features (for instance, their global causality anomalies), as has already been pointed out. Our goal here is primarily to understand the symmetry changes and qualitative phenomena that are possible within CS-modified gravity. The stability for various types of perturbations around G\"{o}del-type metrics has already been investigated in \cite{Barrow:2003ph}, and hereafter, we shall consider only stationary deformations of the G\"{o}del-type metrics in order to help us pick out the properties we are most interested in.

The paper is structured as follows. In section \ref{sec1}, we review the key properties of NDCSMG. Section \ref{sec2} briefly summarizes the main features of G\"{o}del-type metrics and reviews their consistency within the NDCS theory. In section \ref{sec3}, we discuss in detail the perturbation scheme we have adopted and the resulting solutions of the perturbed field equations. Section \ref{sec4} looks at the behavior of the orbital effective potential for planar orbits in these perturbed spacetime backgrounds, and we conclude in section \ref{sec5} with a summary of our overall conclusions. 
   
\section{Review of NDCS modified gravity}
\label{sec1}

In this section we briefly review the main properties of the NDCS modified gravity \cite{Jackiw:2003pm, Alexander:2009tp}. To begin, we write down the action for the model, including a cosmological constant. This is
\begin{equation}
S=\frac{1}{2\kappa}\int d^{4}x\left[\sqrt{-g}(R-2\Lambda)+\frac{1}{4}\phi\,^{*}\!RR\right]+S_{m}(\psi,g_{\mu\nu}),
\label{ac}
\end{equation}
where $S_{m}$ is the matter source action, $\phi$ is a pseudoscalar field with an externally prescribed spacetime dependence, and $^{*}\!RR$ is the Pontryagin density given by
\begin{equation}
^{*}\!RR=\,^{*}\!R^{\lambda\,\,\,\mu\nu}_{\,\,\,\tau}R^{\tau}_{\,\,\,\lambda\mu\nu},
\end{equation}
where the dual of the Riemann curvature tensor is defined by
\begin{equation}
^{*}\!R^{\lambda\,\,\,\mu\nu}_{\,\,\,\tau}=\frac{1}{2}\varepsilon^{\mu\nu\alpha\beta}R^{\tau}_{\,\,\,\lambda\alpha\beta},
\end{equation}
with $\varepsilon^{\mu\nu\alpha\beta}$ being the Levi-Civita symbol. Because the Pontryagin density can be written as a total derivative of the topological current density $K^{\mu}$,
\begin{equation}
^{*}\!RR=2\partial_{\mu}K^{\mu},
\end{equation}
where
\begin{equation}
K^{\mu}=\varepsilon^{\mu\nu\alpha\beta}\left(\Gamma^{\sigma}_{\nu\gamma}\partial_{\alpha}\Gamma^{\gamma}_{\beta\sigma}+\frac{2}{3}\Gamma^{\sigma}_{\nu\gamma}\Gamma^{\gamma}_{\alpha\delta}\Gamma^{\delta}_{\beta\sigma}\right),
\end{equation}
the action (\ref{ac}) can be cast into the following form
\begin{equation}
S=\frac{1}{2\kappa}\int d^{4}x\,\sqrt{-g}\left[(R-2\Lambda)-\frac{1}{2}v_{\mu}\tilde{K}^{\mu}\right]+S_{m}(\psi,g_{\mu\nu}),
\end{equation}
with $\tilde{K}^{\mu}\equiv \frac{K^{\mu}}{\sqrt{-g}}$ being the topological current vector; meanwhile, $v_{\mu}=\partial_{\mu}\phi$ may be interpreted as an axial-vector-valued CS coefficient, which implements CPT violation in the model \cite{Kostelecky:2003fs}. According to this point of view, any physical observable which gets directly coupled to $v_{\mu}$ will experience a CPT symmetry breaking.

The field equation is obtained by varying the action (\ref{ac}) with respect to the metric. This gives
\begin{equation}
G_{\mu\nu}+C_{\mu\nu}-\Lambda g_{\mu\nu}=\kappa T_{\mu\nu},
\label{mod}
\end{equation}  
where $T_{\mu\nu}$ is the stress-energy tensor of the matter sources. In addition to the usual GR terms, the variation of the CS term has given rise to the new ingredient in the modified field equation---the Cotton tensor
\begin{equation}
C^{\mu\nu}=-\frac{1}{2}\left[v_{\alpha}(\epsilon^{\alpha\mu\sigma\tau}\nabla_{\sigma}R^{\nu}_{\,\tau}+\epsilon^{\alpha\nu\sigma\tau}\nabla_{\sigma}R^{\mu}_{\,\tau})+v_{\sigma\tau}(^{*}\!R^{\sigma\mu\tau\nu}+\,^{*}\!R^{\sigma\nu\tau\mu})\right],
\end{equation}
where $v_{\mu\nu}=\nabla_{\mu}v_{\nu}$ and $\epsilon^{\alpha\mu\sigma\tau}\equiv \frac{\varepsilon^{\alpha\mu\sigma\tau}}{\sqrt{-g}}$ is the Levi-Civita tensor. Assuming that the matter content fulfills the energy-momentum conservation conditions, the divergence of eq.\ (\ref{mod}) leads to the well-known Pontryagin constraint 
\begin{equation}
^{*}\!RR=0,
\label{pont}
\end{equation}  	
constraining the space of solutions of the theory.

\section{G\"{o}del-type metrics in NDCS modified gravity}
\label{sec2}

We shall now address the role of G\"{o}del-type metrics in the NDCSMG theory, laying out their most notable features. Before proceeding forward, it is worth stressing that we shall concentrate on the class of G\"{o}del-type metric backgrounds which are homogeneous in space and time (ST-homogeneous). Such metrics are completely characterized by two parameters $m^2$ and $\omega$, as displayed in their line elements \cite{Reb}:
\begin{equation}
ds^2=-\left[dt+H(r)\,d\theta\right]^2+D(r)^{2}\,d\theta^2+dr^2+dz^2,
\end{equation}
with
\begin{eqnarray}
\frac{H'(r)}{D(r)}&=&2\omega,\label{ns1}\\
\frac{D''(r)}{D(r)}&=&m^2, \label{ns2}
\end{eqnarray}
where the prime stands for differentiation with respect to $r$. Eqs.\ (\ref{ns1}) and (\ref{ns2}) are necessary and sufficient conditions for having homogeneity in space and time. The parameter $\omega$ is physically the vorticity, since the G\"{o}del-type metrics describe rotating spacetimes. Depending on the sign of $m^2$, the ST-homogeneous G\"{o}del-type spaces can be separated into three distinct classes:
\begin{itemize}
\item \textit{hyperbolic class}: $m^2>0$, $\omega\neq 0$:
\begin{eqnarray}
 \label{hyper1}H(r)&=&\frac{2\omega}{m^2}\left[\cosh(mr)-1\right],\\
\label{hyper2}D(r)&=&\frac{1}{m}\sinh(mr),
\end{eqnarray}
\item \textit{trigonometric class}: $-\mu^2=m^2<0$, $\omega\neq 0$:
\begin{eqnarray}
 H(r)&=&\frac{2\omega}{\mu^2}\left[1-\cos(\mu r)\right],\\
D(r)&=&\frac{1}{\mu}\sin(\mu r),
\label{trigo}
\end{eqnarray}
\item \textit{linear class}: $m^2=0$, $\omega\neq 0$:
\begin{eqnarray}
 H(r)&=&\omega r^2,\\
D(r)&=&r.
\label{linear}
\end{eqnarray}
\end{itemize}
The special case $m^2=2\omega^2$ of the hyperbolic class corresponds to the famous original G\"{o}del metric~\cite{Godel}. More general ST-homogeneous G\"{o}del-type spaces present different isometry groups, according to the relations between their metric parameters. For instance, the range $m^2<4\omega^2$ admits $G_5$ as the isometry group, but on the other hand, $m^2=4\omega^2$ admits $G_{7}$ as the isometry group.

A remarkable property of ST-homogeneous G\"{o}del-type spaces (or, more generally, of cyl\-in\-dri\-cal\-ly-symmetric spacetimes \cite{Bronnikov:2019clf}) is the presence of CTCs, running along circular trajectories given by $C=\lbrace(t,r,\theta,z)\,| \, t,\, r,\,\mbox{and}\, z\, \mbox{constant}; \theta \in [0, 2\pi]\rbrace$, where the function $G(r)=D(r)^{2}-H(r)^{2}$ assumes a nonpositive value. For the hyperbolic class of spacetimes, the critical radius $r_{c}$, defined as the limiting radius allowing for the existence of these circular CTCs, is given by      
\begin{equation}
\sinh^2\bigg(\frac{m r_{c}}{2}\bigg)=\bigg(\frac{4\omega^2}{m^2}-1 \bigg)^{-1}.
\label{rc}
\end{equation}  
Note that for $m^2\geq 4\omega^2$ the existence of CTCs is entirely circumvented, regardless of the value of the $r$ coordinate. On the other hand, for $m^2<4\omega^2$ the existence of CTCs is unavoidable for $r<r_c$. In much the same way, the other (trigonometric and linear) classes present similar CTCs. (See \cite{Reb} for a detailed discussion.)    

Having outlined the most relevant properties of the ST-homogeneous G\"{o}del-type metrics, we are now able to discuss them within the NDCSMG. The authors in \cite{Porfirio:2016nzr} have shown that ST-homogeneous G\"{o}del-type metrics are solutions of the modified field equations (\ref{mod}), fulfilling the Pontryagin constraint for physically-motivated matter sources, namely: a perfect fluid; a material scalar field with linear dependence of the $z$-coordinate, i.e., $\psi(z)=s(z-z_{\psi})$, where $s$ and $z_{\psi}$ are constants; and a sourceless electromagnetic field whose electric and magnetic components lie along the $z$-direction (see \cite{Porfirio:2016nzr} to get the explicit expressions for them). The CS pseudoscalar field supporting these kinds of solution is of the $z$-dependent form $\phi(z)=b(z-z_0)$, where $b$ and $z_0$ are constants; as a consequence, its gradient, the CS coefficient $v_{\mu}$, lies along the cosmic rotation axis. Another interesting property of this solution without any analogy in GR is that the gradient $v_{\mu}$, couples to the vorticity vector $\omega_{\mu}=\delta^{z}_{\mu}\omega$, providing thereby a new effective parameter $k=b\omega$ in the field equation. This new coupling allows for completely causal solutions ($m^2\geq 4\omega^2$) even in the presence of all the aforementioned matter content types, depending on the sign of $k$ as shown in \cite{Porfirio:2016nzr}.        

In the next section we shall deal with first-order perturbations to ST-homogeneous G\"{o}del-type metrics and their $\phi$ parameters which nonetheless leave the matter sources unperturbed. Our aim will be to check whether one can generate analytical solutions for the first-order field equations that entail breaking of the cylindrical symmetry or the translational invariance along the $z$-direction. Obviously, either of these would lead to the breakdown of the ST-homogeneity and then would probably affect the causality properties.

\section{Perturbative scheme for non-homogeneous G\"{o}del-type metrics}
\label{sec3}

In this section we shall examine the possibility of breaking the ST-homogeneity for G\"{o}del-type metrics within NDCSMG. To do that, we shall introduce a perturbative approach, which will be discussed in detail below. Then we can straightforwardly substitute the perturbed metric into the field equations to find the full solutions up to first order in the perturbation parameter. 

\subsection{The perturbative scheme}
It has been shown in \cite{Porfirio:2016nzr} that ST-homogeneous G\"{o}del-type metrics are solutions of NDCSMG, since they can satisfy the Pontryagin constraint---although they do not reduce to GR solutions, because of the fact that the Cotton tensor is generally nontrivial (except for the special class with $m^2=4\omega^2$). 

In this paper, we shall restrict our attention to stationary perturbations, since the G\"{o}del-type metrics are themselves stationary. This implies that the perturbation functions depends only on the coordinates $r$, $\theta$ and $z$. By virtue of explicit dependences on $\theta$ and $z$, the axial symmetry and translational invariance along the $z$-direction may be broken. Thus, the resulting metrics are no longer ST homogeneous. We label the ST-homogeneous G\"{o}del-type background metric $g_{\mu\nu}^{(0)}$, while the first-order perturbation to the metric is denoted $\xi g_{\mu\nu}^{(1)}$. Using this notation, we can write down the perturbed metric as follows
\begin{equation}
g_{\mu\nu}=g_{\mu\nu}^{(0)}+\xi g_{\mu\nu}^{(1)}+\mathcal{O}(\xi^2),
\label{pertm}
\end{equation}    
so that $\xi$ is an explicit perturbation parameter, and we shall consider perturbations in all our equations only up to first order in $\xi$.  

To proceed further, let us expand eq.\ (\ref{pertm}) explicitly. Doing this, we have
\begin{eqnarray}
\nonumber ds^2&=&-\left\{\left[1+\xi h_{0}(r,\theta,z)\right]dt+\left[1+\xi h_{1}(r,\theta,z)\right]H(r)\,d\theta\right\}^2+\left[1+\xi h_{2}(r,\theta,z)\right]D(r)^{2}\,d\theta^{2} \\
& & +\left[1+\xi h_{3}(r,\theta,z)\right]dr^{2}+\left[1+\xi h_{4}(r,\theta,z)\right]dz^{2},
\label{pertm1}
\end{eqnarray}
where the $h_i(r,\theta,z)$ are the functions which characterize the first-order metric $g_{\mu\nu}^{(1)}$. As can be seen from the perturbed metric, we are restricting our analysis to metric perturbations for which $g_{\mu\nu}^{(1)}=0$ for all off-diagonal components in this coordinate basis except for $g_{t\theta}^{(1)}$. This mirrors the structure of the background metric $g_{\mu\nu}^{(0)}$, whose only nonzero components are $g_{t\theta}^{(0)}$ and the other $g_{\mu\nu}^{(0)}$ with $\mu=\nu$. Moreover, we consider a more general CS field
\begin{equation}
\phi=\phi^{(0)}+\xi \phi^{(1)}(r,\theta,z),
\end{equation}
where $\phi^{(0)}$ is the linear background field discussed above. The perturbation $\phi^{(1)}(r,\theta,z)$ will potentially contribute to Cotton tensor only at the first order in $\xi$.

\subsection{Pontryagin constraint}

Any solution of NDCSMG must satisfy the Pontryagin constraint. This striking ingredient will impose constraints on the perturbed metric functions. To check that the constraints hold, we must evaluate $^{*}\!RR$ for eq.\ (\ref{pertm1}). According to a straightforward calculation, the Pontryagin constraint up to first order in $\xi$ becomes
\begin{eqnarray}
0 & = & (4\omega^{2}-m^{2})\left[-2\omega D(r)\frac{\partial}{\partial z}h_{0}(r,\theta,z) +\omega D(r)
\frac{\partial}{\partial z}h_{3}(r,\theta,z)-2 \omega D(r)
\frac{\partial}{\partial z}h_{1}(r,\theta,z)\right. \nonumber\\
&& +\,\omega D(r)\frac{\partial}{\partial z}h_{2}(r,\theta,z)+H(r)\frac{\partial^{2}}{\partial z\,
\partial r}h_{0}(r,\theta,z)-\left.H(r)\frac{\partial^{2}}{\partial z\,\partial r}h_{1}(r,\theta,z)\right].
\label{eq-Pont}
\end{eqnarray}
For the sake of convenience, eq. (\ref{eq-Pont}) can be cast in the form
\begin{equation}
H(r)\frac{\partial^{2}}{\partial z\,\partial r}P(r,\theta,z)-\omega D(r)
\frac{\partial}{\partial z}Q(r,\theta,z)=0
\label{HD}
\end{equation}
(for $m^2\neq 4\omega^2$) by defining the new quantities
\begin{eqnarray}
P(r,\theta,z)&\equiv& h_{0}(r,\theta,z)-h_{1}(r,\theta,z)\\
Q(r,\theta,z)&\equiv& 2\left[h_{0}(r,\theta,z)+h_{1}(r,\theta,z)\right]-\left[h_{2}(r,\theta,z)+h_{3}(r,\theta,z)\right].
\end{eqnarray}

The simplest solution of eq.\ (\ref{HD}) corresponds to $P=Q=0$, thus enforcing a relationship among the perturbed metric functions, namely,
\begin{equation}
h_{0}(r,\theta,z)=h_{1}(r,\theta,z)=\frac{1}{4}\left[h_{2}(r,\theta,z)+h_{3}(r,\theta,z)\right].
\label{const}
\end{equation}
As we shall see, such a constraint may be used to simplify the field equations to allow for an analytical solution.

\subsection{Field equations}

We are here interested in the first-order perturbed field equations, since we already know that the background ones are self-consistent. We shall thus employ the aforementioned perturbation scheme; that is, only the metric and $\phi$ will be perturbed, whilst the matter sources remain unaltered. Furthermore, we will take eq.\ (\ref{const}) to hold, in order to ensure that the Pontryagin constraint is satisfied. Following these assumptions, the first-order field equations clearly reduce to
\begin{equation}
G_{\mu\nu}^{(1)}+C_{\mu\nu}^{(1)}=0,
\end{equation}
with the perturbed Einstein and Cotton tensors.

To begin solving the first-order field equations, it is useful to start with the specific component
$G_{tz}^{(1)}+C_{tz}^{(1)}=0$, which gives
\begin{equation}
0=\frac{\partial^{3}}{\partial z\,\partial \theta\,\partial r}h_{2}(r,\theta,z)-
\frac{\partial^{3}}{\partial z\,\partial \theta\,\partial r}h_{3}(r,\theta,z)
+ \left[\frac{\partial^{2}}{\partial z\,\partial\theta}h_{2}(r,\theta,z)-
\frac{\partial^{2}}{\partial z\,\partial\theta}h_{3}(r,\theta,z)\right]
\frac{d}{dr}\ln D(r),
\end{equation}
with a solution
\begin{equation}
h_{2}(r,\theta,z)-h_{3}(r,\theta,z)=f(z,r)+g(\theta,r)+\frac{h(\theta,z)}{D(r)},
\label{eqh}
\end{equation}
that ties $h_{{2}}\left( r,\theta,z \right)$ to $h_{{3}}\left( r,\theta,z \right)$. Moreover, for the $rz$-component, $\theta z$-component, and $zz$-com\-po\-nent of the field equation, we have, respectively,
\begin{eqnarray}
0 & = & \frac{3}{2}\frac{\partial^{2}}{\partial z\,\partial r}h_{2}(r,\theta,z)+
\frac{1}{2}\frac{\partial^{2}}{\partial z\,\partial r}h_{3}(r,\theta,z)
+ \left[\frac{\partial}{\partial z}h_{2}(r,\theta,z)-
\frac{\partial}{\partial z}h_{3}(r,\theta,z)\right]\frac{d}{dr}\ln D(r), \label{eq1} \\
0 & = & \left[\frac{\partial^{2}}{\partial z\,\partial\theta}h_{3}(r,\theta,z)-
\frac{\partial^{2}}{\partial z\,\partial\theta}h_{2}(r,\theta,z)\right]
\frac{d}{dr}\ln D(r)+
\left[3\frac{\partial^{2}}{\partial z\,\partial\theta}h_{3}(r,\theta,z)\right. \nonumber\\
& & +\left.\frac{\partial^{2}}{\partial z\,\partial\theta}h_{2}(r,\theta,z)\right]\frac{D(r)}{bH(r)}+
\frac{\partial^{3}}{\partial z\,\partial \theta\,\partial r}h_{3}(r,\theta,z)-
\frac{\partial^{3}}{\partial z\,\partial \theta\,\partial r}h_{2}(r,\theta,z), \label{eq2} \\
0 & = & 3\frac{\partial^{2}}{\partial r^{2}}h_{3}(r,\theta,z)+
\frac{\partial^{2}}{\partial r^{2}}h_{2}(r,\theta,z)-16\left[h_{2}(r,\theta,z)+h_{3}(r,\theta,z)+h_{4}(r,\theta,z)\right]
\omega^{2}+\left[14h_{3}(r,\theta,z)+\right. \nonumber\\
&+&\left. 18h_{2}(r,\theta,z)+16h_{4}(r,\theta,z)\right]+\left[5\frac{\partial}{\partial r}h_{2}(r,\theta,z)-\frac{\partial}{\partial r}h_{3}(r,\theta,z)\right]\frac{d}{dr}\ln D(r)+\nonumber\\
&+&\left[\frac{\partial^{2}}{\partial\theta^{2}}h_{2}(r,\theta,z)+
3\frac{\partial^{2}}{\partial\theta^{2}}h_{3}(r,\theta,z)\right]\frac{1}{D(r)^{2}}. \label{eq3}
\end{eqnarray}
Substituting eq.\ (\ref{eqh}) into eqs.\ (\ref{eq1}--\ref{eq3}), we obtain 
\begin{eqnarray} 
\label{h2}h_{2}(r,\theta,z)&=&\frac{Z(z)}{D(r)^2},\\
\label{h3}h_{3}(r,\theta,z)&=&-\frac{Z(z)}{D(r)^2},\\
h_{0}(r,\theta,z)&=&h_{1}(r,\theta,z)=h_{4}(r,\theta,z)=0,
\end{eqnarray}
where $Z(z)$ is an arbitrary function of $z$.
 
The perturbation $\phi^{(1)}$ will appear in the remaining nontrivial field equations; these are the $tt$-, $tr$-, $t\theta$-, $rr$-, $r\theta$-, and $z\theta$-components. Upon substituting eqs.\ (\ref{h2}) and (\ref{h3}) into these components, we arrive at five additional equations,
\begin{eqnarray}
0 & = & (4\omega^{2}-m^{2})\frac{\partial}{\partial z}\phi^{(1)}(r,\theta,z), \label{11} \\
0 & = & (4\omega^{2}-m^{2})\frac{\partial^{2}}{\partial z\,\partial\theta}\phi^{(1)}(r,\theta,z), \label{12} \\
0 & = & 2\omega(4\omega^{2}-m^{2})\left[\frac{\partial}{\partial z}\phi^{(1)}(r,\theta,z)\right]H(r)
+\frac{1}{2}(4\omega^{2}-m^{2})\left[\frac{\partial^{2}}{\partial z\,\partial\theta}\phi^{(1)}(r,\theta,z)\right]
D(r), \label{13} \\
0 & = & (4\omega^{2}-m^{2})\frac{\partial}{\partial z}\phi^{(1)}(r,\theta,z)
-\frac{\frac{1}{2\omega}(1-b\omega)\frac{d^{2}}{dz^{2}}Z(z)+[(4\omega^{2}-m^{2})b+\omega]Z(z)}{D(r)^{2}},
\label{22} \\
0 & = & \omega[(4\omega^{2}-m^{2})b+\omega]Z(z)+(4\omega^{2}-m^{2})\left\{\omega
\left[\frac{\partial}{\partial z}\phi^{(1)}(r,\theta,z)\right]\left[2H(r)^{2}+D(r)^{2}\right]
\right. \nonumber\\
& & +\left.\left[\frac{\partial^{2}}{\partial z\,\partial r}\phi^{(1)}(r,\theta,z)\right]D(r)H(r)\right\}-
\frac{1}{2}(1-b\omega)\frac{d^{2}}{dz^{2}}Z(z). \label{23}
\end{eqnarray}
It is noteworthy that there are only five equations because the $tr$- and $r\theta$-components coincide. The system of partial differential equations, eqs.\ (\ref{11}--\ref{23}), has an exact solution given by 
\begin{eqnarray}
\label{Zz}Z(z) & = & C_{1}\sin(\alpha z) +C_{2}\cos(\alpha z),\\
\label{phi}\phi^{(1)}(r,\theta,z) & = & F_{1}(r,\theta), 
\end{eqnarray} 
where
\begin{equation}
\alpha=\sqrt{\frac{2\omega\left[(4\omega^{2}-m^{2})b+\omega\right]}{b\omega-1}},
\end{equation}
$C_{{1}}$ and $C_{{2}}$ are integration constants, and $F_{1}(r,\theta)$ is an arbitrary function of $r$ and $\theta$. It is perhaps unsurprising that $\phi^{(1)}(r,\theta,z)$ is thus unrestricted, since the pseudoscalar $\phi$ field has no independent dynamics of its own. However, the quantity $\alpha$ must be real, in order to avoid exponentially runaway behavior as a function of $z$; such behavior would invalidate the perturbative approach. The condition that $\alpha$ be real then imposes constraints on the allowed ranges for some of the parameters. If $k=b\omega>1$, then there are solutions in the $m^2<4\omega^2$ region of the parameter space, a region in which there is non-causal behavior. Alternatively, if $b\omega<1$, then the solutions are found in the completely causal region, $m^2>4\omega^2$. This allows us to conclude that the perturbed solutions may be either causal and non-causal, in either case in accordance with the behavior of the corresponding background solutions.

Having found the metric functions by inserting eqs.\ (\ref{Zz}) and (\ref{phi}) into eqs.\ (\ref{h2}) and (\ref{h3}), we finally obtain the nonzero first-order-perturbed metric components
\begin{eqnarray}
g_{\theta\theta}^{(1)}&=&C_{1}\sin(\alpha z) +C_{2}\cos (\alpha z), \\
g_{rr}^{(1)}&=&-\frac{1}{D(r)^2}\left[C_{1}\sin(\alpha z) +C_{2}\cos(\alpha z)\right].
\end{eqnarray} 
It should be noted that a $g_{rr}^{(1)}$ of this form has an apparent singularity at $r=0$ for each of the three classes of solution discussion in section~\ref{sec2}. However, it is actually just a coordinate singularity, since an evaluation of the Kretschmann scalar within the first-order perturbation scheme yields a finite value at $r=0$. In the other limit, $g_{rr}^{(1)}$ falls off to zero as $r\rightarrow\infty$, leaving only the the component $g_{\theta\theta}^{(1)}$ to survive in this asymptotic limit.

\subsection{Properties of the solutions}

It follows from the solution (\ref{phi}) that the gradient vector $v_{\mu}$ can take a much more general form in our perturbed theory than was previously possible with just the background $\phi^{(0)}$. The richer structure we have uncovered in the perturbed theory includes the possibility of having a purely spacelike $v_{\mu}$ with all three spatial components being nontrivial,
\begin{equation}
v_{\mu}=\left[0, \xi\frac{\partial F_{1}}{\partial r}, \xi\frac{1}{r}\frac{\partial F_{1}}{\partial \theta}, b\right].
\end{equation}
This pseudovector breaks the axial symmetry and translational invariance. Note, however, that it has constant norm up to first order in $\xi$: $v^{\mu}v_{\mu}=b^2 +\mathcal{O}(\xi^2)$, which is just the value in the unperturbed limit.

For the spacetime metric, the translational invariance along the $z$-axis is spoiled at the first order in $\xi$, and thus so is the overall spatial homogeneity; however, the metric retains its axial symmetry. In contrast, the CS field $\phi$ can exhibit a nontrivial $\theta$-dependence, thus breaking the overall axial symmetry; physical observables coupled to the $v_{\mu}=\partial_{\mu}\phi$ vector are generally sensitive to a nontrivial azimuthal $\theta$-dependence as well.

\section{Planar Geodesics}
\label{sec4}

In this section, we shall investigate the geodesic trajectories of particles moving in the perturbed background. In order to do this, it is first valuable to get information about the Killing vectors associated with the spacetime. As just pointed out in section~\ref{sec3}, the translational symmetry of the metric has been broken, which implies that there remain only two Killing vectors, $\partial_t$ and $\partial_\theta$ (unlike the unperturbed metric, which possesses three Killing vectors). We will focus specifically on planar geodesics---that is, those ones restricted to a fixed-$z$ plane.
 
Our starting point for finding the planar geodesics is writing down the Lagrangian for a point particle \cite{Wald}
\begin{equation}
\label{eql}\mathcal{L}=g_{\mu\nu}\dot{x}^{\mu}\dot{x}^{\nu},
\end{equation}
where the dot stands for a derivative with respect to an affine parameter $\lambda$. The scaled values $\mathcal{L}=-1$, $0$, and $1$ denote timelike, null, and spacelike geodesics, respectively. The one-form velocity $U$ of the particle can be decomposed into
\begin{equation}
U=-E\,dt + U_{r}\,dr+L\,d\theta, 
\end{equation} 
with its radial component given by
\begin{equation}
U_{r}=\left[1+\frac{\xi C_{{2}}}{D(r)^{2}}\right]\dot{r},
\end{equation}
where we have picked the plane corresponding to $z=0$, without any meaningful loss of generality. The constants of motion $E$ and $L$ are associated with the Killing vectors $\partial_{t}$ and $\partial_{\theta}$; their explicit forms are
\begin{eqnarray}
\label{eqa}E&=&\dot{t}+H(r)\dot{\theta},\\
\label{eqb}L&=&\left[D(r)^{2}-H(r)^{2}-\xi C_2\right]\dot{\theta}-H(r)\dot{t}.
\end{eqnarray}

Now substituting eqs.\ (\ref{eqa}) and (\ref{eqb}) into eq.\ (\ref{eql}), we obtain the differential equation for the radial coordinate along the geodesics in terms of the conserved quantities,
\begin{equation}
\frac{\dot{r}^{2}}{2}=\frac{E^2}{2}-\frac{\xi E^2 C_2}{2 D(r)^{2}}-\frac{1}{2}\left[\frac{E H(r)+L}{D(r)}\right]^2+\frac{1}{2}\mathcal{L}\left[1-\frac{\xi C_2}{D(r)^{2}}\right].
\end{equation}
 This equation describes the classical motion of a particle with unit mass, with energy $\mathcal{E}=\frac{E^2}{2}$, and in the presence of the effective potential 
\begin{equation}
\label{Veff}V_{{\it eff}}=\frac{\xi E^2 C_2}{2 D(r)^{2}}+\frac{1}{2}\left[\frac{E H(r)+L}{D(r)}\right]^2-\frac{1}{2}\mathcal{L}\left[1-\frac{\xi C_2}{D(r)^{2}}\right].
\end{equation}
Therefore, the radial equation can be rewritten as
\begin{equation}
\frac{\dot{r}^{2}}{2}=\mathcal{E}-V_{{\it eff}}.
\end{equation}
 The modifications stemming from the perturbation parameter $\xi$ in the effective potential can be explicitly seen by rewriting eq.\ (\ref{Veff}) as $V_{{\it eff}}=V_{{\it eff}}^{(0)}+\frac{\xi}{2}\left[\frac{E^2 C_2}{D(r)^{2}}+\frac{\mathcal{L}C_2}{D(r)^{2}}\right] + \mathcal{O}(\xi^2)$, where $V_{{\it eff}}^{(0)}$ is the effective potential of the unperturbed metric. Therefore, the effects of the first-order perturbations  will depend upon which class the G\"{o}del-type metric we are dealing with belongs to. Hereafter, we shall restrict our analysis to the hyperbolic class, described by eqs.\ (\ref{hyper1}) and (\ref{hyper2}), for which the effective potential takes the following form
\begin{equation}
V_{{\it eff}}=\frac{\xi m^2 E^2 C_2}{2 \sinh^{2}(mr)}+\frac{1}{2}\left\{\frac{2E\omega[\cosh(mr)-1]+L m^2}{m\sinh(mr)}\right\}^2-\frac{1}{2}\mathcal{L}\left[1-\frac{\xi m^2 C_2}{\sinh^{2}(mr)}\right].
\label{Ve2}
\end{equation}

\begin{figure}
		\subfloat[]{\includegraphics[width=0.5\textwidth]{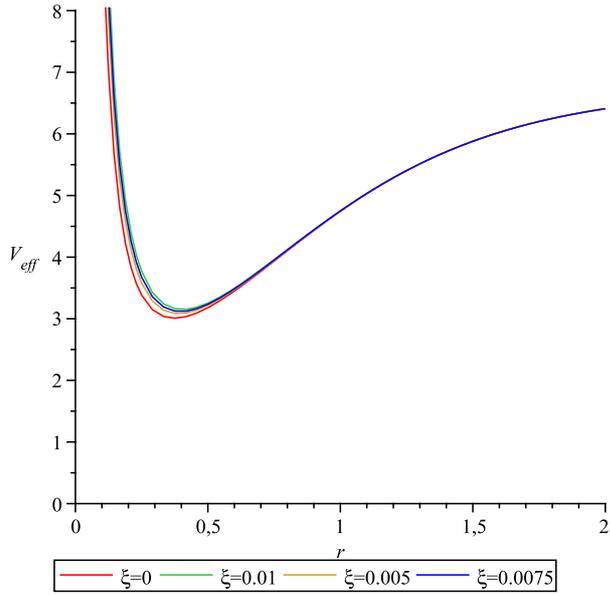}}
	\subfloat[Zooming in on the minimum region of (a).]{\includegraphics[width=0.5\textwidth]{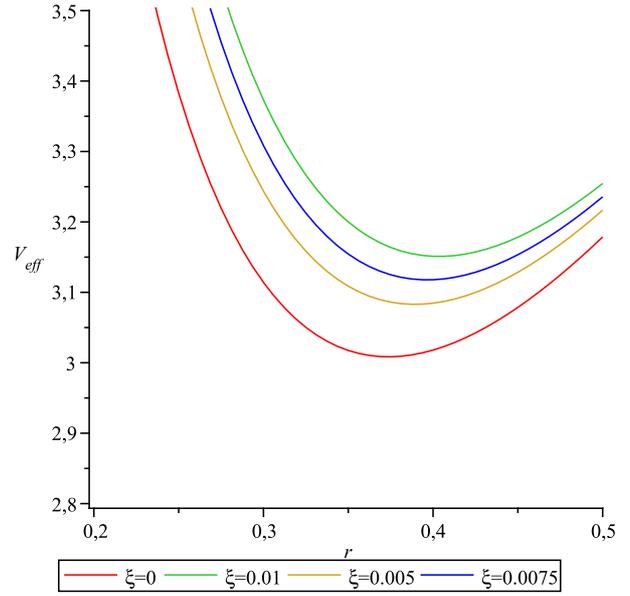}}
	\caption{Effective potential $V_{{\it eff}}$ for several values of $\xi$, assuming other parameter values of $\mathcal{L}=-1$, $m=2$, $\omega=\sqrt{2}$, $C_{2}=1$, $L=0.4$, and $E=2.5$. The curves for $\xi=0$ correspond to the unperturbed metric.}
	\label{fig:graphpert11}
\end{figure}

\begin{figure}
		\subfloat[]{\includegraphics[width=0.5\textwidth]{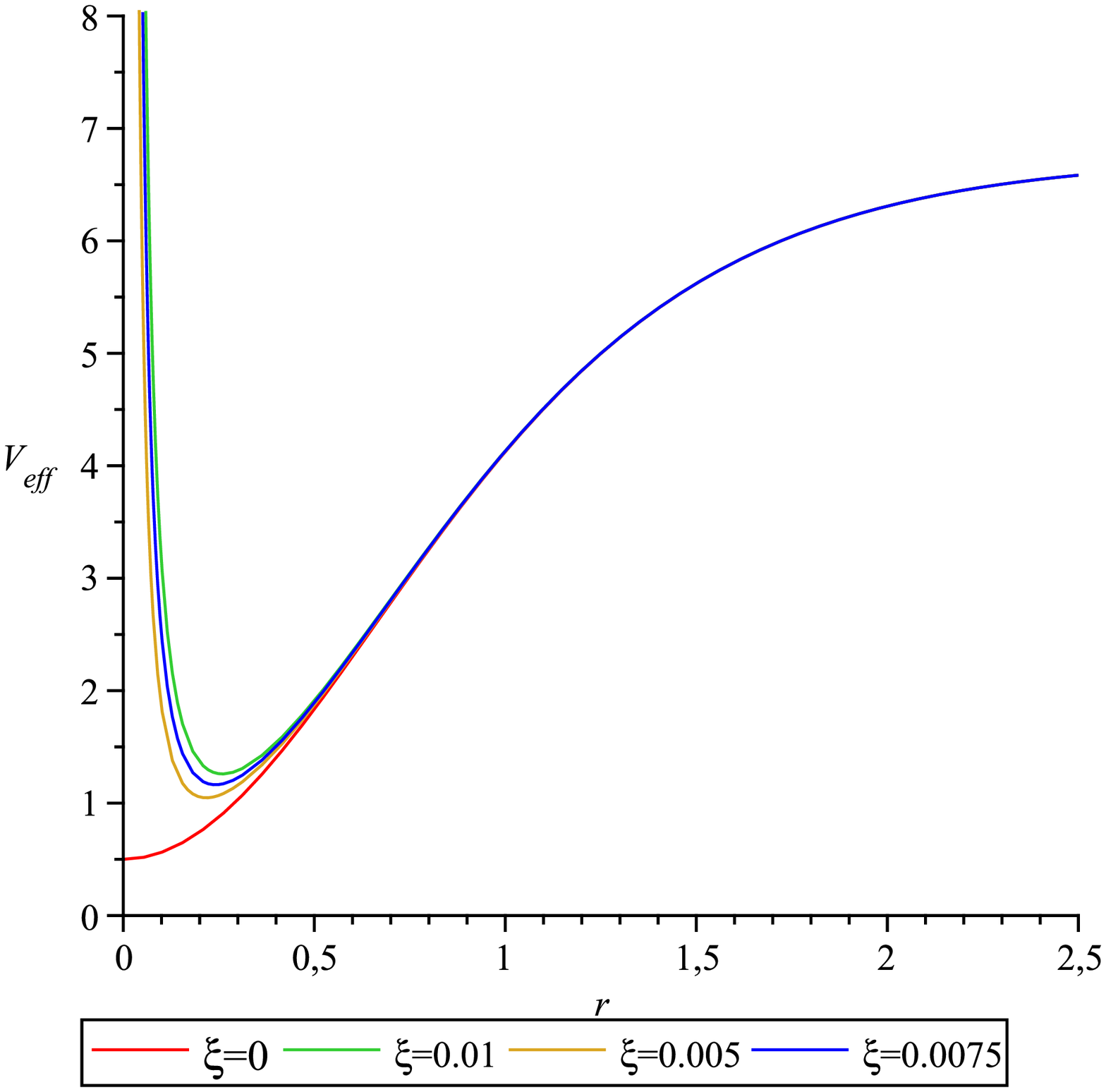}}
	\subfloat[Zooming in on the minimum region of (a).]{\includegraphics[width=0.5\textwidth]{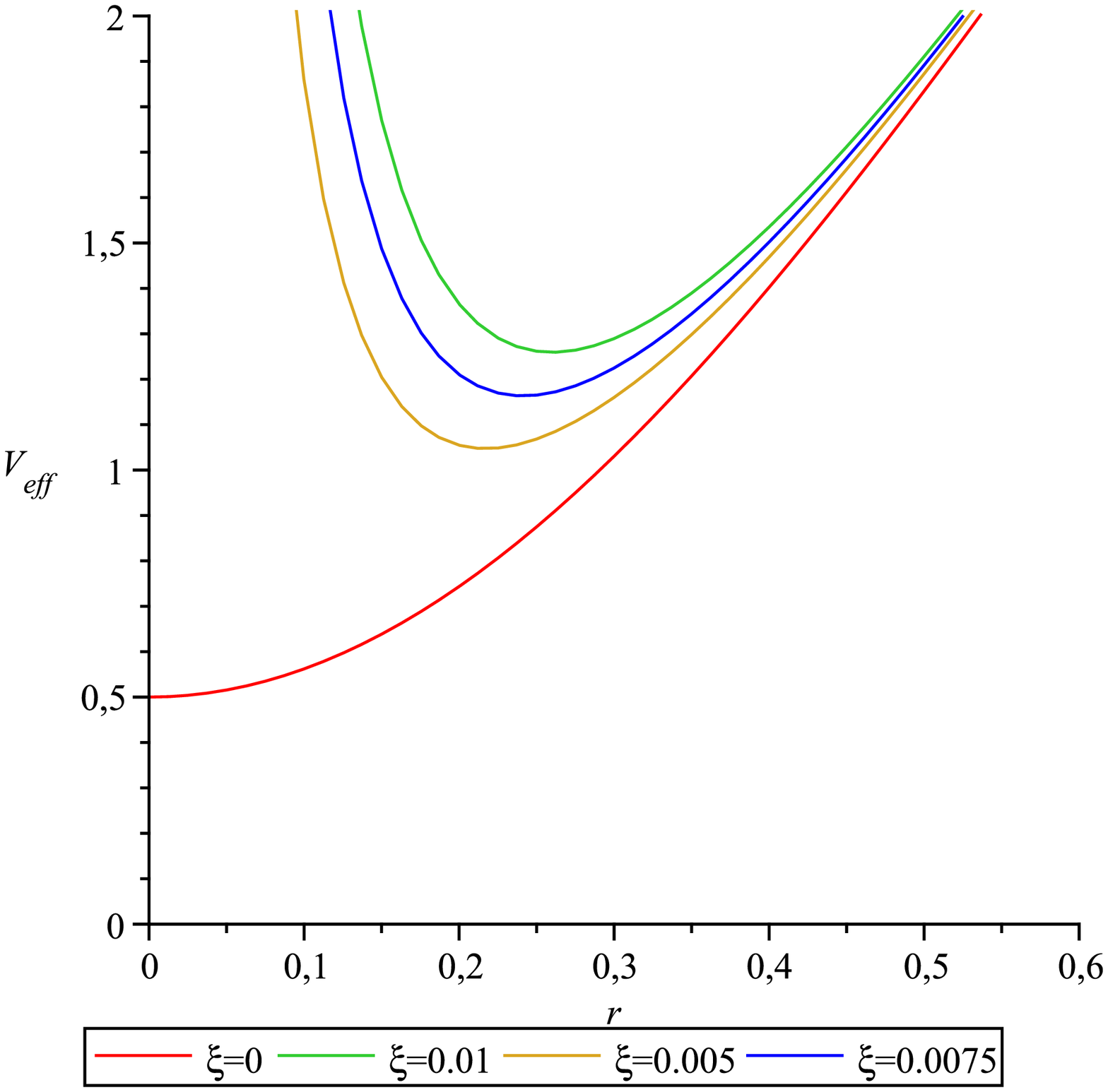}}
	\caption{$V_{{\it eff}}$ for several values of $\xi$, assuming other parameter values of $\mathcal{L}=-1$, $m=2$, $\omega=\sqrt{2}$, $C_{2}=1$, $L=0$, and $E=2.5$.}
	\label{fig:graphpert22}
\end{figure}

\begin{figure}
		\subfloat[]{\includegraphics[width=0.5\textwidth]{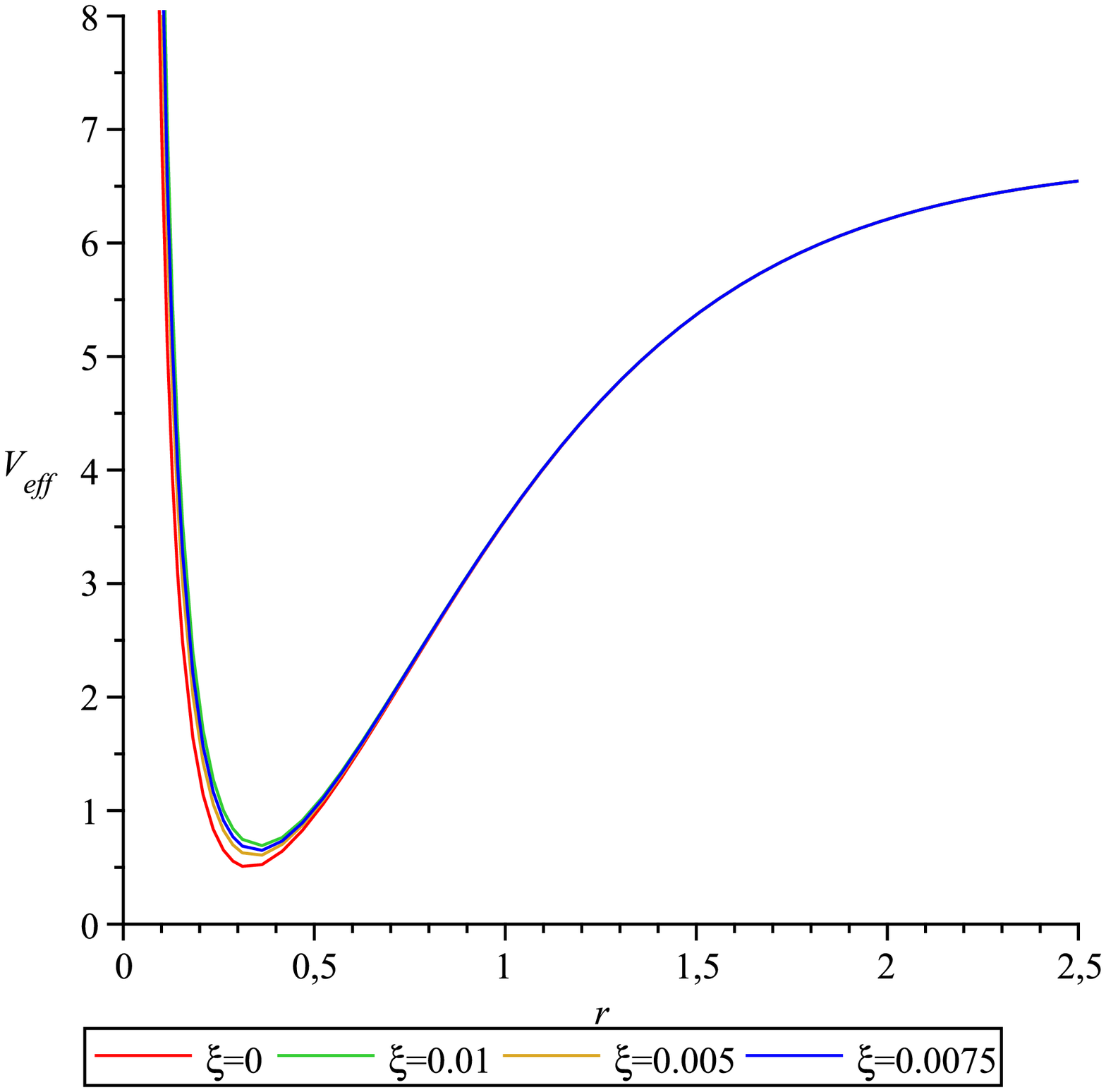}}
	\subfloat[Zooming in on the minimum region of (a).]{\includegraphics[width=0.5\textwidth]{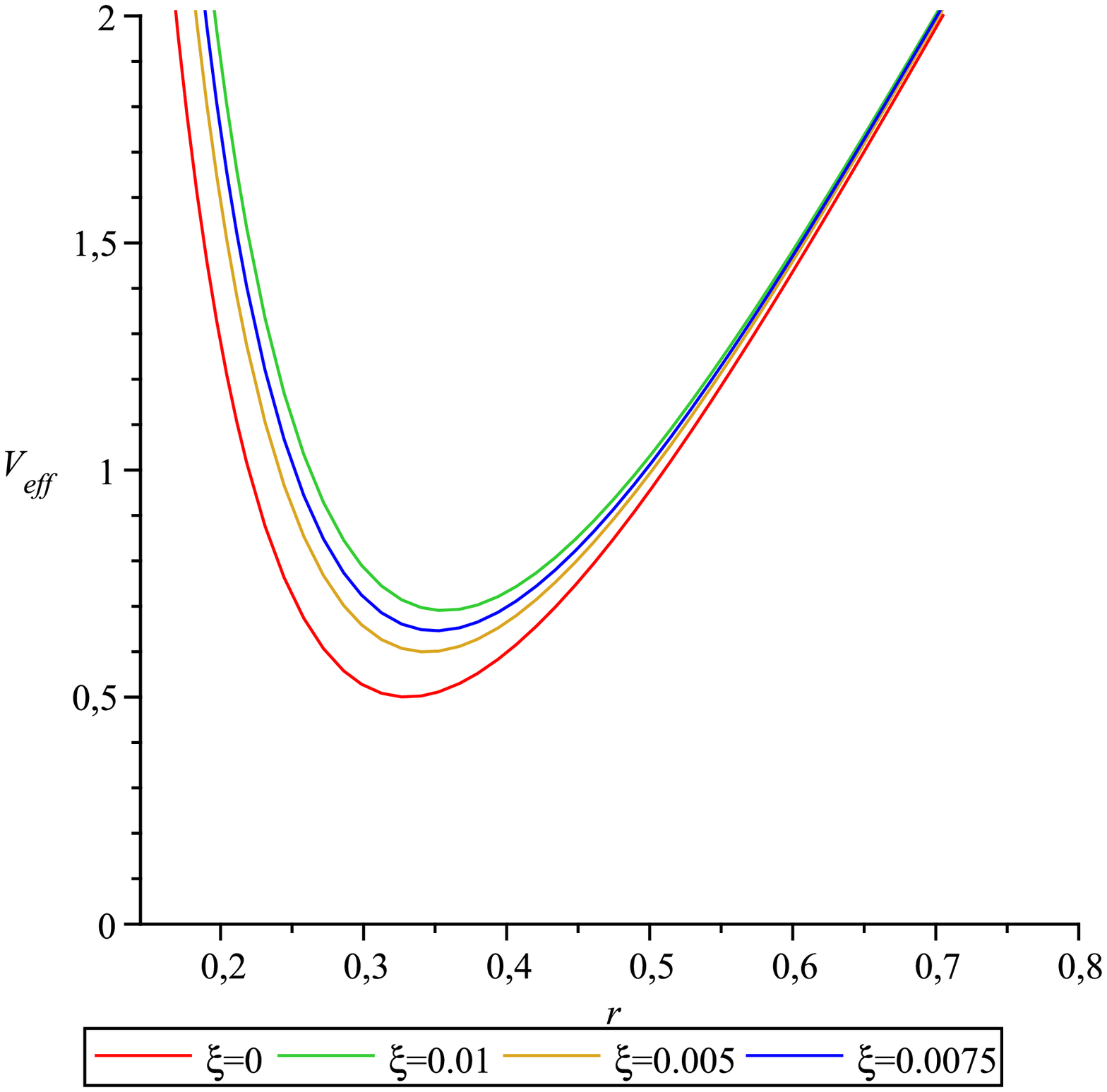}}
	\caption{$V_{{\it eff}}$ for several values of $\xi$, assuming other parameter values of $\mathcal{L}=-1$,  $m=2$, $\omega=\sqrt{2}$, $C_{2}=1$, $L=-0.4$, and $E=2.5$.}
	\label{fig:graphpert33}
\end{figure}

Now we are able to examine the motion for both massive particles (timelike geodesics) and photons (null geodesics). We know that the allowed radial motions are restricted to the range $\mathcal{E}-V_{{\it eff}}>0$, with the turning points being at $\mathcal{E}=V_{{\it eff}}$. The extremum of the effective potential $\frac{d V_{{\it eff}}}{dr}=0$ corresponds to circular orbits of constant radius, and they are stable when $\frac{d^{2} V_{{\it eff}}}{d r^2}>0$, which means that a particle tends to return to its initial radial position after small a displacement from the equilibrium radius. 

\begin{figure}
		\subfloat[]{\includegraphics[width=0.5\textwidth]{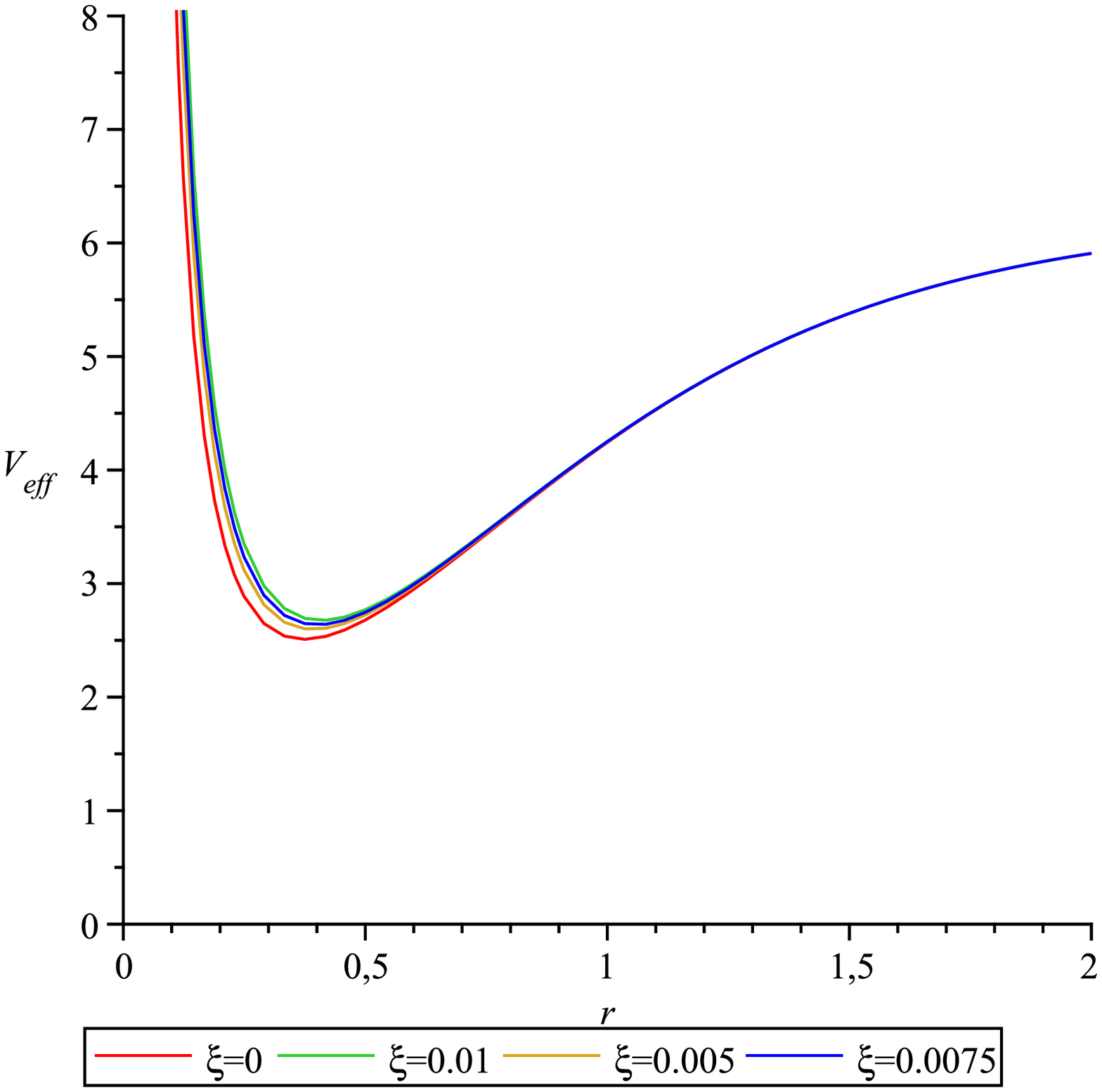}}
	\subfloat[Zooming in on the minimum region of (a).]{\includegraphics[width=0.5\textwidth]{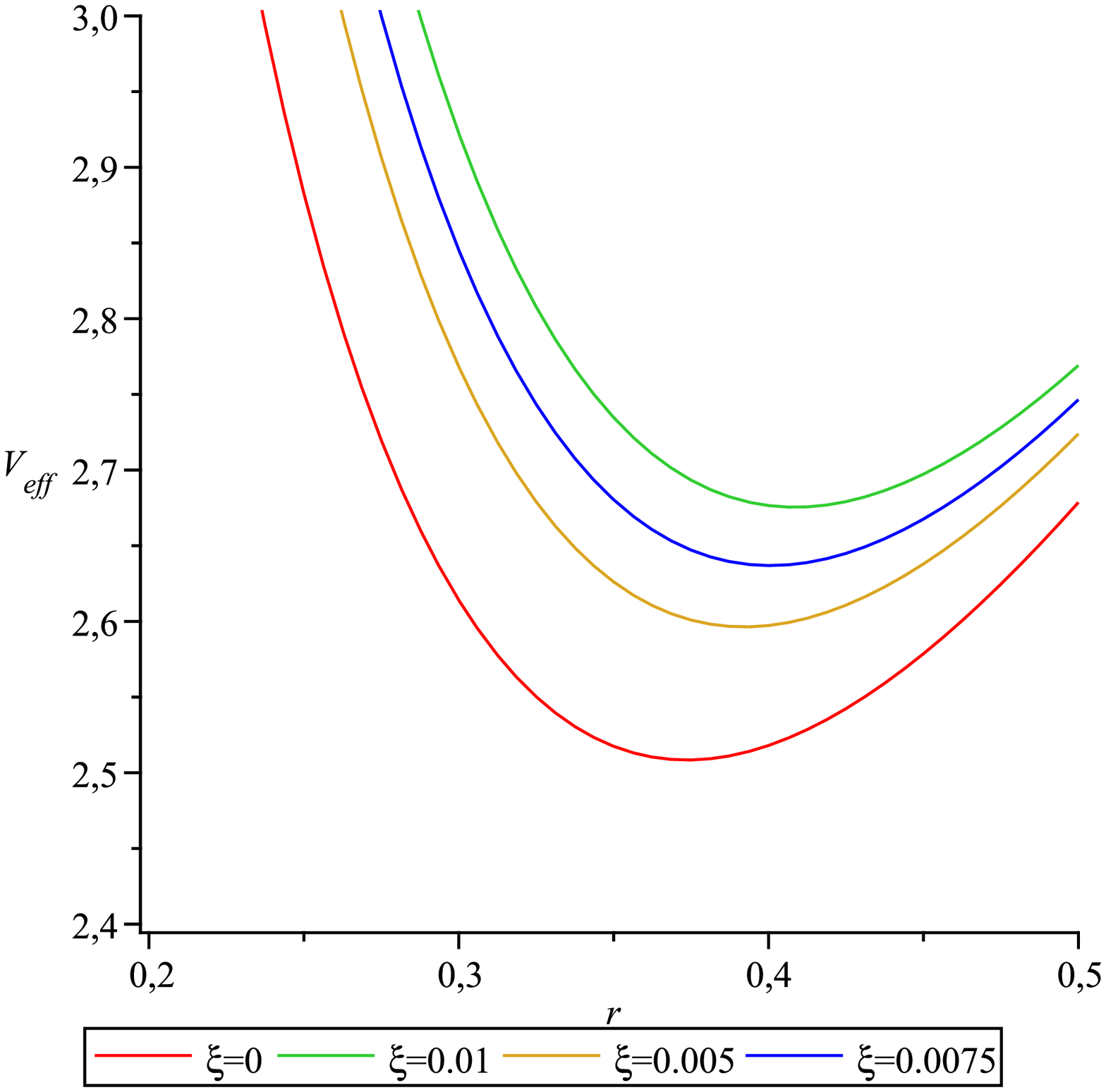}}
	\caption{$V_{{\it eff}}$ for several values of $\xi$, assuming other parameter values of $\mathcal{L}=0$, $m=2$, $\omega=\sqrt{2}$, $C_{2}=1$, $L=0.4$ and $E=2.5$.}
	\label{fig:graphpert44}
\end{figure}

\begin{figure}
		\subfloat[]{\includegraphics[width=0.5\textwidth]{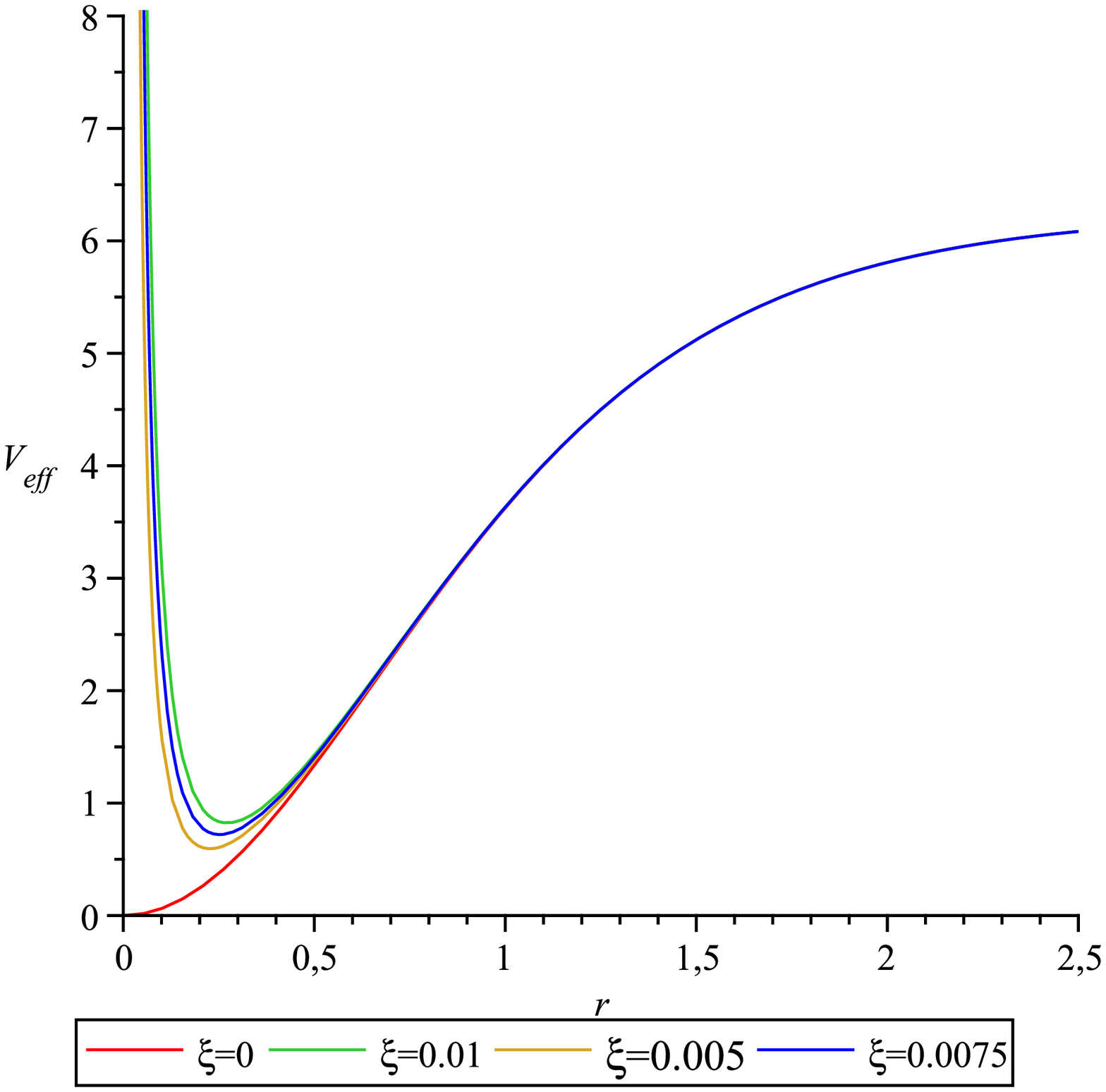}}
	\subfloat[Zooming in on the minimum region of (a).]{\includegraphics[width=0.5\textwidth]{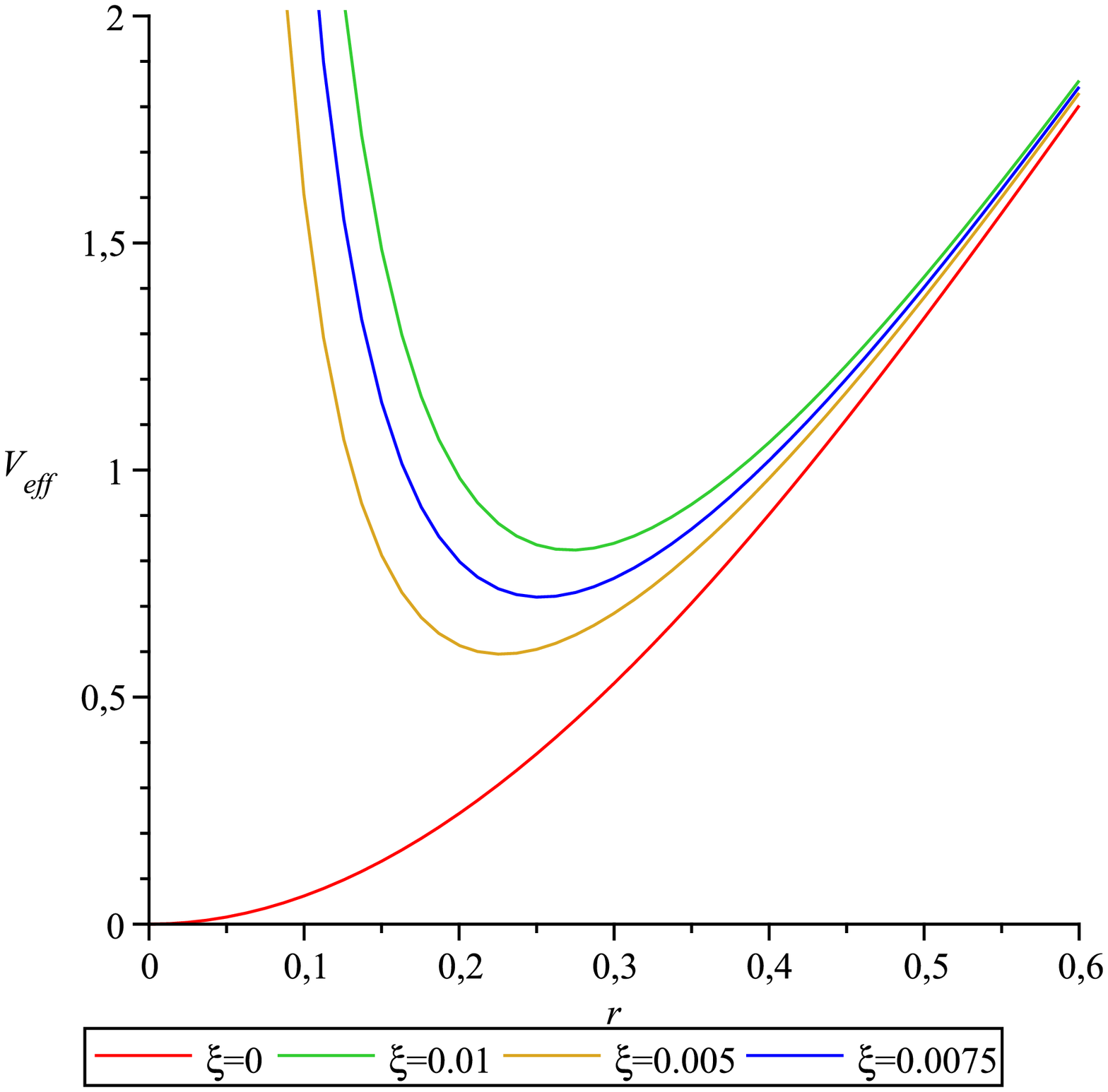}}
	\caption{$V_{{\it eff}}$ for several values of $\xi$, assuming other parameter values of $\mathcal{L}=0$, $m=2$, $\omega=\sqrt{2}$, $C_{2}=1$, $L=0$ and $E=2.5$.}
	\label{fig:graphpert55}
\end{figure}

\begin{figure}
		\subfloat[]{\includegraphics[width=0.5\textwidth]{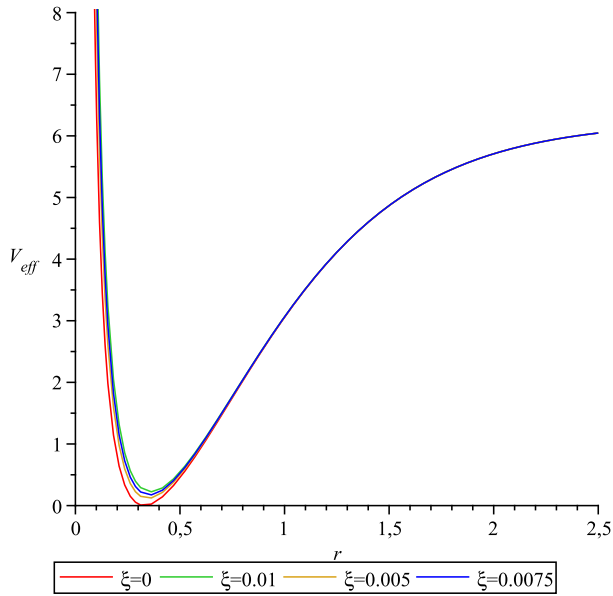}}
	\subfloat[Zooming in on the minimum region of (a).]{\includegraphics[width=0.5\textwidth]{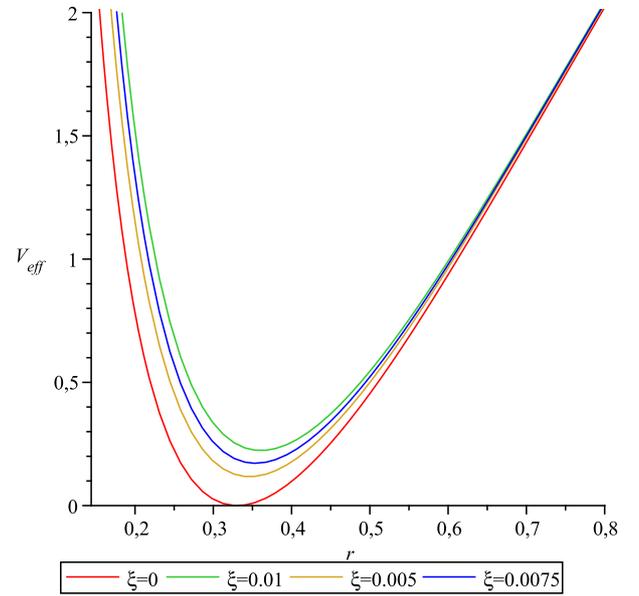}}
	\caption{$V_{{\it eff}}$ for several values of $\xi$, assuming other parameter values of $\mathcal{L}=0$, $m=2$, $\omega=\sqrt{2}$, $C_{2}=1$, $L=-0.4$ and $E=2.5$.}
	\label{fig:graphpert66}
\end{figure}

From eq.\ (\ref{Ve2}), we can discriminate three different types of behavior for the effective potential, depending on the values of $L$, namely $L>0$, $L<0$, and $L=0$. Figs. (\ref{fig:graphpert11}--\ref{fig:graphpert66}) display plots of $V_{{\it eff}}(r)$ for those three different cases, first for massive particles (timelike geodesics, $\mathcal{L}=-1$) and then for photons (null geodesics, $\mathcal{L}=0$). For the $\mathcal{L}=-1$ case, figs.\ (\ref{fig:graphpert11}--\ref{fig:graphpert33}) display the shape of the effective potential. In particular, fig.\ (\ref{fig:graphpert11}) shows the shape of the effective potential for $L>0$ for a variety of timelike geodesic orbits, corresponding to a number of small positive values of the perturbation parameter $\xi$. It is noteworthy that, as can readily be seen in fig.\ (\ref{fig:graphpert11}), the most significant effects due to a $\xi\neq0$ occur only in the neighborhood of the minimum of $V_{{\it eff}}$. It is also apparent that, as $\xi$ grows, the minimum of the effective potential moves toward larger values of $r$.
    
The shape of the effective potential for $L=0$ is depicted in fig.\ (\ref{fig:graphpert22}). We note a peculiar phenomenon that does not appear in the $L\neq 0$ cases. As in the case $L>0$ case, an increasing $\xi$ shifts the minimum of $V_{{\it eff}}$ to the right. However, for sufficiently small values of $r$, the values of the effective potential are markedly different from their unperturbed ($\xi=0$) values, regardless of how small $\xi$ is. Instead of reaching its minimum at $r=0$, when $\xi\neq 0$ there is an additional inner turning point, and at smaller values of $r$, $V_{{\it eff}}$ increases rapidly. This singular behavior signals a breakdown of the perturbation scheme for the effective potential in this region   
		
Fig.\ (\ref{fig:graphpert33}) displays the behavior of the effective potential for $L<0$. In this case, the effects of $\xi>0$ are quite similar to what they were for $L>0$, and there is again no qualitatively anomalous behavior at small $r$.
		
We now turn our attention to null geodesics, which correspond to $\mathcal{L}=0$ in eq.\ (\ref{Ve2}). The effective potentials are depicted in figs.\ (\ref{fig:graphpert44}--\ref{fig:graphpert66}). Each of the plots shows similar behavior to what was found in the massive particle cases, and the qualitative conclusions regarding the effects of the perturbation parameter on the effective potential are the same as for the $\mathcal{L}=-1$ orbits.

\section{Summary and conclusions}
\label{sec5}

In this paper, we have dealt with first-order perturbations on top of the ST-homogeneous G\"{o}del-type metric solutions of the NDCS modified gravity theory, departing from the background solution found in \cite{Porfirio:2016nzr} that had a spacelike CS vector $v_{\mu}$ pointing along the $z$-direction about which the local frames are rotating. We considered, to first order in the parameter $\xi$, temporally stationary perturbations that preserved the property of the G\"{o}del-type metrics that there be only a single nonzero off-diagonal (frame-dragging) metric component. However, the perturbed theory was allowed to break the axial symmetry and the translational invariance along the $z$-direction. The first-order perturbation the CS pseudoscalar field $\phi$ was thus also permitted to have dependences on $r$, $\theta$ and $z$, although the Pontryagin constraint leads to nontrivial relations among the perturbed metric functions, which do not involve the perturbations to $\phi$.

Taking into account the Pontryagin constraint, it turns out that solutions of the first-order field equations can be found analytically. The solutions provide a spacelike vector breaking of the axial symmetry at the first-order perturbation level. The first-order perturbations to the metric also break the translational symmetry along the $z$-axis and the homogeneity of the spacetime. This is a remarkable result, since it is the first solution in which the NDCS modification to the theory breaks the the ST-homogeneity conditions.

With regard to the global properties of the deformed metric, we have shown that the perturbed metric functions we found do not affect the large-scale causality properties of the spacetime. In other words, the existence of CTCs depends only on the background (unperturbed) metric as it was in the absence of the small deformation. This is interesting but not unexpected, since it seems unreasonable for small perturbations to affect such globally defined properties of ST-homogeneous G\"{o}del-type metrics as CTCs. On the other hand, we have also shown that local properties, such as the motions of massive and massless particles, are nontrivially affected by the perturbations; the perturbation parameter $\xi$ modifies the effective potential for radial motion along planar geodesics, with increasing values of $\xi$ generally pushing the orbits to be larger than their values in GR ($\xi=0$).

One natural continuation of this study would be to consider the same questions in the dynamical CS gravity theory. It would also be  interesting also to consider the effects at a higher orders in perturbation theory, and we plan to examine both of these questions further in forthcoming work.

\vspace{.5cm}
   
\textbf{Acknowledgments}  
This work was partially supported by Conselho
Nacional de Desenvolvimento Cient\'{\i}fico e Tecnol\'{o}gico (CNPq). The work by A. Yu. P. has been supported by the CNPq project No. 301562/2019-9. P. J. P. would like to thank the Brazilian agency CAPES for financial support.

\end{document}